\documentclass{jfm}
\usepackage{Paper}

\shorttitle{}
\shortauthor{N. Nair and A. Goza}

\title[State and reduced state estimation with deep learning]{Leveraging reduced-order models for state estimation using deep learning}

\author{Nirmal J. Nair\aff{1}
  \corresp{\email{njn2@illinois.edu}}
  \and Andres Goza\aff{1}}

\affiliation{\aff{1}Department of Aerospace Engineering, University of Illinois at Urbana-Champaign, IL 61801, USA}

\newcommand{\RR}{\ensuremath{\mathbb{R}}} 
 
\newcommand{\map}{\ensuremath{\mathcal{G}}} 

\newcommand{\Ltwo}{\ensuremath{\ell_{2}}} 

\newcommand{\state}{\ensuremath{\bm{w}}}
\newcommand{\stateapprox}{\ensuremath{\bm{w}_r}}
\newcommand{\timet}{\ensuremath{t}}
\newcommand{\timetmax}{\ensuremath{t_{max}}}
\newcommand{\parameter}{\ensuremath{\bm{\mu}}}
\newcommand{\dynamics}{\ensuremath{\bm{f}}}
\newcommand{\parameteroned}{\ensuremath{\mu}}
\newcommand{\parameteronedpred}{\ensuremath{\mu^*}}

\newcommand{\parameterpred}{\ensuremath{\bm{\mu}^*}}
\newcommand{\parameterpredn}[1]{\ensuremath{\bm{\mu}^{*}_{#1}}}
\newcommand{\parametersampled}{\ensuremath{\bm{\mu}^{s}}}
\newcommand{\parametersampledn}[1]{\ensuremath{\bm{\mu}^{s}_{#1}}}
\newcommand{\parameteronedsampled}{\ensuremath{\mu^s}}
\newcommand{\parameteronedsampledn}[1]{\ensuremath{\mu^{s}_{i}}}

\newcommand{\statesize}{\ensuremath{N_{w}}}
\newcommand{\parametersize}{\ensuremath{N_{d}}}
\newcommand{\nmodes}{\ensuremath{N_{k}}}
\newcommand{\nsensors}{\ensuremath{N_{s}}}
\newcommand{\nsnapshots}{\ensuremath{M}}
\newcommand{\nlayers}{\ensuremath{N_{l}}}
\newcommand{\layersize}[1]{\ensuremath{l_{#1}}}
\newcommand{\nparams}{\ensuremath{N_{p}}}
\newcommand{\nparamspred}{\ensuremath{N^*_{p}}}
\newcommand{\ntimes}{\ensuremath{N_{t}}}

\newcommand{\ntrain}{\ensuremath{N_{train}}}
\newcommand{\nvalid}{\ensuremath{N_{valid}}}
\newcommand{\ntest}{\ensuremath{N_{test}}}
\newcommand{\ntrainsnapshots}{\ensuremath{M_{train}}}
\newcommand{\nvalidsnapshots}{\ensuremath{M_{valid}}}
\newcommand{\ntestsnapshots}{\ensuremath{M_{test}}}

\newcommand{\basis}{\ensuremath{\bm{\Phi}}}
\newcommand{\trialbasis}{\ensuremath{\bm{\Psi}}}
\newcommand{\gc}{\ensuremath{\bm{a}}}
\newcommand{\gcdata}{\ensuremath{\bm{A}}}
\newcommand{\sensor}{\ensuremath{\bm{s}}}
\newcommand{\sensordata}{\ensuremath{\bm{S}}}

\newcommand{\samplingmatrix}{\ensuremath{\bm{C}}}

\newcommand{\linearmap}{\ensuremath{\bm{G}}}
\newcommand{\nonlinearmap}{\ensuremath{\bm{g}}}

\newcommand{\weights}{\ensuremath{\bm{\theta}}}
\newcommand{\hiddenlayer}[1]{\ensuremath{\bm{h}_{#1}}}
\newcommand{\weightmat}[1]{\ensuremath{\bm{\Theta}_{#1}}}
\newcommand{\activation}{\ensuremath{\sigma}}

\newcommand{\reynolds}{\ensuremath{Re}}

\newcommand{\dx}{\ensuremath{\Delta x}}
\newcommand{\dt}{\ensuremath{\Delta t}}

\newcommand{\timeinstant}{\ensuremath{n}}
\newcommand{\staten}[1]{\ensuremath{\state^{#1}}}
\newcommand{\statemean}{\ensuremath{\bar{\state}}}
\newcommand{\stateapproxn}[1]{\ensuremath{\stateapprox^{#1}}}
\newcommand{\sensorn}[1]{\ensuremath{\sensor^{#1}}}
\newcommand{\gcn}[1]{\ensuremath{\gc^{#1}}}

\newcommand{\gcdummyn}[1]{\ensuremath{\hat{\gc}^{#1}}}
\newcommand{\linearmapdummy}{\ensuremath{\hat{\linearmap}}}
\newcommand{\weightsdummy}{\ensuremath{\hat{\weights}}}
\newcommand{\dummyinput}{\ensuremath{\bm{\xi}}}

\begin{document}

\maketitle

\begin{abstract}
State estimation is key to both analyzing physical mechanisms and enabling real-time control of fluid flows. A common estimation approach is to relate sensor measurements to a \emph{reduced state} governed by a reduced-order model (ROM). (When desired, the full state can be recovered via the ROM). Current methods in this category nearly always use a linear model to relate the sensor data to the reduced state, which often leads to restrictions on sensor locations and has inherent limitations in representing the generally nonlinear relationship between the measurements and reduced state. We propose an alternative methodology where a neural network architecture is used to learn this nonlinear relationship. Neural network  is a natural choice for this estimation problem, as a physical interpretation of the reduced state-sensor measurement relationship is rarely obvious. The proposed estimation framework is agnostic to the ROM employed, and can be incorporated into any choice of ROMs derived on a linear subspace (\emph{e.g.}, proper orthogonal decomposition) or a nonlinear manifold. The proposed approach is demonstrated on a two-dimensional model problem of separated flow around a flat plate, and is found to outperform common linear estimation alternatives.

\end{abstract}



\section{Introduction}

In fluid dynamics, the goal of state estimation (SE) is to accurately estimate the instantaneous flow field using a set of limited sensor measurements. Achieving this goal can provide insights into key physics and facilitate the prediction and control of flows in various engineering applications. In many problems where state estimation is of interest, a reduced-order model (ROM) of the high-dimensional system is also typically available. Accordingly, a class of SE strategies that leverage this low-order representation have emerged---that is, estimation is done on a \emph{reduced state} obtained from the ROM (the full state can be recovered via the ROM when desired). In this article, we focus on such methods which are particularly promising for real-time control applications\footnote{Many successful SE methods do not rely on a low-order representation of the flow state. Examples include sparse identification using SINDy \citep{loiseau2018sparse}, identifying problem-specific parameters via an ensemble Kalman filter \citep{darakananda2018data} or convolutional autoencoder \citep{hou2019machine}, or state estimation using a shallow decoder \citep{erichson2019shallow}. However, it may be computationally prohibitive to integrate these SE approaches into ROMs due to an intermediate step that involves the high-dimensional fluid state.}.


In most model order reduction approaches, the dynamics of the high-dimensional state are projected onto a low-dimensional linear subspace. A number of bases for this subspace have been developed; \emph{e.g.}, proper orthogonal decomposition (POD) \citep{Lumley1967TheSO}, dynamic mode decomposition \citep{schmid2010dynamic} and balanced POD \citep{willcox2002balanced}. More recently, nonlinear ROMs have been developed that utilize local bases instead of a global basis \citep{amsallem2012nonlinear}, or a global nonlinear manifold constructed using autoencoders from deep learning \citep{lee2019model, otto2019linearly}. 

Estimation of the reduced state derived from ROMs can be broadly divided into two categories: intrusive and non-intrusive. Intrusive SE models such as Kalman filtering \citep{kalman1960new} and particle filtering \citep{gordon1993novel} rely on an observer dynamical system to predict the state (which is later updated based on observed data). These data-assimilation approaches have been coupled with POD-based ROMs on various flow problems \citep{kikuchi2015assessment, tu2013integration}. On the other hand, non-intrusive methods are model-free and can be further classified into library and non-library based approaches.

In library based approaches, the sensor measurements are approximated with the same library that is used for the ROM (\emph{e.g.}, obtained from POD modes \citep{bright2013compressive} or the training data itself \citep{callaham2019robust}). The resulting optimization problem can be solved in the $\ell_1$ norm to promote sparsity \citep{candes2006near}. Alternatively, the reduced-state can be estimated in the $\ell_2$ norm, termed gappy-POD \citep{everson1995karhunen}. To overcome  ill-conditioning and overfitting in this $\ell_2$ setting, sensor locations can be chosen through greedy \citep{clark2018greedy}, optimal \citep{brunton2013optimal} or sparse \citep{sargsyan2015nonlinear} sensor placement algorithms, which can outperform $\ell_1$-based approaches \citep{manohar2018data}. However, the need for problem-specific sensor locations in this estimation framework limits its flexibility.

Non-library based approaches, on the other hand, provide an empirically determined map between the measurements and reduced state. This alleviates restrictions on sensor locations and ill-conditioning inherent to library-based methods. One example is linear stochastic estimation (LSE), which provides a linear map through an $\ell_2$ minimization of available data \citep{adrian1975role}. Although traditional LSE relates sensor measurements to the high-dimensional state, recent variants estimate the reduced state \citep{taylor2004towards, podvin2018combining}. Quadratic stochastic estimation \citep{murray2007modified} provides a specific nonlinear extension to LSE. However, for complex fluid flow problems the nonlinear relationship between the sensor measurements and the reduced state is generally unknown, and a more flexible framework is necessary.

In this work, we model this nonlinear relationship using neural networks. This approach allows for a lower number of sensors and greater flexibility in sensor locations compared with its linear counterparts. We demonstrate the efficacy of our approach on a two-dimensional model problem of separated flow past a flat plate, and compare results to those obtained via gappy-POD and LSE. 
While our results on the model problem are obtained using a POD-based ROM, we emphasize that our formulation is agnostic to the ROM, and can be incorporated into either linear or nonlinear ROMs. 


\section{State-estimation: ROM-based framework and prior work}

\subsection{ROM-based state estimation framework}

Consider the dynamical system resulting from the semi-discretization of partial differential equations such as the Navier-Stokes equations:
\begin{equation}
\dot{\state} = \dynamics(\state, \timet; \parameter), \quad \state(\timet_{\timeinstant};\parameter)=\staten{\timeinstant}(\parameter)
\label{fom}
\end{equation}
where $\state(\timet; \parameter) \in \RR^{\statesize}$ represents the high-dimensional state that depends on time $\timet \in [0,$\timetmax$]$ and a vector of $\parametersize$ parameters $\parameter \in \RR^{\parametersize}$. The nonlinear function $\dynamics(\state,\timet;\parameter)$ governs the dynamics of the state $\state(\timet;\parameter)$. Provided an estimate of an instantaneous state $\staten{\timeinstant}(\parameter)$ at an arbitrary time  $\timet_{\timeinstant}$, the initial value problem \eqref{fom} can be used to determine $\staten{\timeinstant+i}(\parameter)$ for $i=1,2,\ldots$. We refer to Eq. \eqref{fom} as the full-order model (FOM).

We consider the scenario where a reduced-order model (ROM) of \eqref{fom} is available. In this case, the high-dimensional state $\state(\timet; \parameter)$ is approximated on a low-dimensional manifold as 
\begin{equation}
\state(\timet;\parameter) \approx \stateapprox(\timet;\parameter) = \basis( \gc(\timet;\parameter))
\label{romapprox}
\end{equation}
where $\basis: \RR^{\nmodes} \mapsto \RR^{\statesize}$ denotes the nonlinear manifold, $\gc(\timet;\parameter) \in \RR^{\nmodes}$ is the reduced state on this manifold, and $\nmodes \ll \statesize$ is the dimension of the reduced state. To facilitate a clean presentation of the ROM, we assume that $\basis(\gc)$ is continuously differentiable such that $\bm{\Upsilon}(\dot{\gc})=\dot{(\basis (\gc))}$ for some $\bm{\Upsilon}: \RR^{\nmodes} \mapsto \RR^{\statesize}$. Substituting the ROM approximation \eqref{romapprox} in Eq. \eqref{fom}, and projecting the resulting equation onto a test manifold $\bm{\Lambda}: \RR^{\statesize} \mapsto \RR^{\nmodes}$ such that $\bm{\Lambda} \bm{\Upsilon}$ is injective, yields
\begin{equation} 
\dot{\gc} = \trialbasis(\dynamics(\basis (\gc),\timet;\parameter)); \ \ \gc(\timet_n;\parameter)=\gcn{n}(\parameter)
\label{rom}
\end{equation}
where $\trialbasis(\cdot)=(\bm{\Lambda} \bm{\Upsilon})^{-1} \circ \bm{\Lambda}(\cdot)$ and $\gcn{n}(\parameter)$ is the initial condition at time instant $\timet_{\timeinstant}$ for the new initial value problem \eqref{rom}. In the case of Galerkin projection where $\basis$ and $\bm{\Lambda}=\basis^T$ are linear and orthogonal, $\trialbasis = \basis^T$. 

\begin{figure}
\centering
\includegraphics[scale=1]{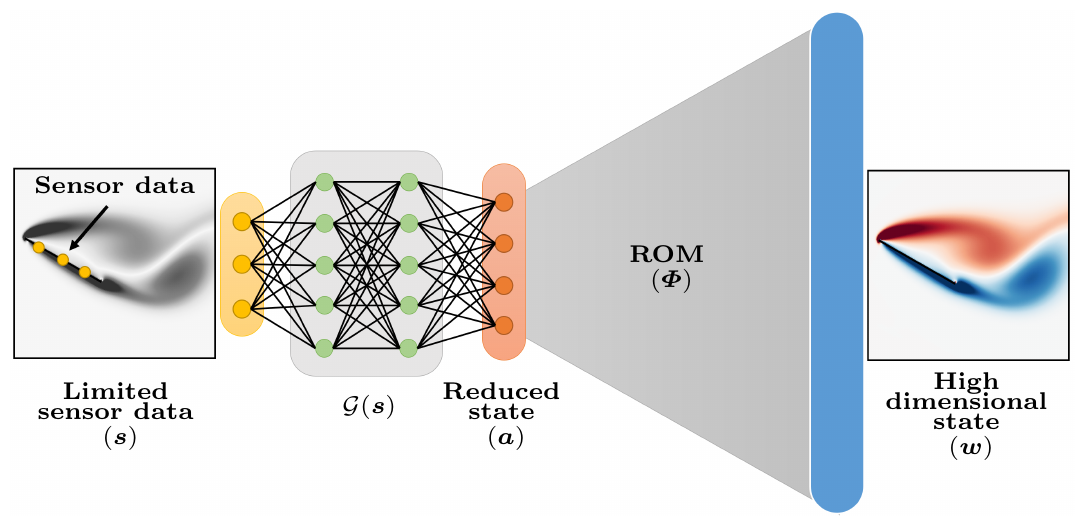}
\caption{Schematic of the ROM-based state estimation framework, which employs a low-dimensional representation $\basis$ of the high-dimensional state $\staten{n}$. $\map$ maps sensor measurements $\sensorn{n}$ to the reduced state $\gcn{n}$. In the proposed approach described in Sec. \ref{DSE}, $\map$ is nonlinear and constructed from a neural network (as pictured). For traditional linear estimation models described in Sec \ref{prior}, $\map$ represents a real-valued matrix and the pictured neural network is no longer valid.}
\label{schematic}
\end{figure}

Now, the original SE goal of estimating the instantaneous high-dimensional state $\staten{\timeinstant}(\parameter)$ reduces to estimating the lower dimensional state  $\gcn{\timeinstant}(\parameter)$. That is, the SE problem amounts to identifying a map $\map : \RR^{\nsensors} \mapsto \RR^{\nmodes}$ between the sensor measurements and the reduced-state such that $\gcn{n} = \map(\sensorn{n})$,
%
%
where $\sensorn{n}(\parameter) \in \RR^{\nsensors}$ denotes the sensor measurements at time instant $\timeinstant$ and $\nsensors$ is the number of sensors in the flow-field. A schematic of the ROM-based SE framework described here is displayed in Fig. \ref{schematic}.

\subsection{Prior work: linear estimation models}
\label{prior}

The traditional approach of identifying the map $\map$ is given by gappy-POD \citep{everson1995karhunen}. In this approach, $\basis$ is restricted to be linear and the sensors directly measure the high-dimensional state at $\nsensors<\statesize$ flow locations, such that $\sensorn{n}(\parameter) = \samplingmatrix \staten{n}(\parameter)  \approx \samplingmatrix \basis \gcn{n}(\parameter)$.
%
%
%
%
The matrix $\samplingmatrix \in \RR^{\nsensors \times \statesize} $ contains one at measurement locations and zero at all other locations.
The reduced state $\gcn{n}$ is obtained by the minimization problem, 
\begin{equation}
\gcn{n} = \underset{\gcdummyn{n}}{\text{arg min}} \| \sensorn{n} - \samplingmatrix \basis \gcdummyn{n} \|_2^2
\label{gpod2}
\end{equation}
The solution to Eq. \eqref{gpod2} is analytically provided by the Moore-Penrose pseudo inverse of $\samplingmatrix \basis$, resulting in the linear map $\map$ to be defined as $\gcn{n} = \map(\sensorn{n}) = (\samplingmatrix \basis)^+ \sensorn{n}$.

Linear stochastic estimation (LSE) can be considered as a generalization of gappy-POD where the sensor measurements are not restricted to lie in the span of the basis of the high-dimensional state. In other words, the linear operator $(\samplingmatrix \basis)^+$ can be replaced by a more general matrix $\linearmap \in \RR^{\nmodes \times \nsensors}$---that is, $\map$ is represented via $\gcn{n} = \map(\sensorn{n}) = \linearmap \sensorn{n}$.
%
%
In LSE, $\linearmap$ is determined from data via the optimization problem,
\begin{equation}
\linearmap = \underset{\linearmapdummy}{\text{arg min}} \| \sensordata - \linearmapdummy \gcdata \|_2^2
\label{lse1}
\end{equation}
where $\sensordata \in \RR^{\nsensors \times \nsnapshots}$ and $\gcdata \in \RR^{\nmodes \times \nsnapshots}$ are snapshot matrices of sensor measurements and reduced states, respectively, consisting of $\nsnapshots$ snapshots. 
The solution to Eq. \eqref{lse1} is analytically obtained as $\linearmap = \sensordata \gcdata^+$.
%
%


\subsubsection{Drawbacks of linear estimation models}
\label{lsedrawbacks}

In gappy-POD, the sensor locations encoded in $\samplingmatrix$ can significantly influence the condition number of $\samplingmatrix \basis$. In particular, sensor locations are required to coincide with regions where the columns of $\basis$ have significant nonzero and distinct values. Sensor locations that do not satisfy this property lead to an inaccurate estimation of the reduced state, $\gcn{n} = (\samplingmatrix \basis)^+ \sensorn{n}$. Furthermore, choosing more library elements than sensors, $\nmodes>\nsensors$, can result in overfitting. These limitations can be resolved by selecting optimal sensor locations that improve the condition number of $\samplingmatrix \basis$ \citep{manohar2018data}  and/or incorporating regularization in Eq. \eqref{gpod2} to mitigate overfitting. However, the need to select specific sensor locations reduces the flexibility of gappy-POD.

Unlike gappy-POD, LSE is significantly more robust to sensor locations. However, it linearly models a (generally nontrivial) nonlinear relationship between the sensor measurements and the reduced state. Therefore, this approach is limited by the rank of $\linearmap$, which is at best $\text{rank}(\linearmap) \leq \text{min}(\nmodes,\nsensors)$. Estimation performance of LSE can be improved by increasing the number of sensors, $\nsensors$, though this is not always possible depending on the given application. We propose a method for more robustly recovering the reduced state by learning a nonlinear relationship using a neural network.



\section{Proposed approach: deep state estimation (DSE)}
\label{DSE}

In our proposed approach, the map $\map$ is further generalized to nonlinearly relate sensor measurements to the reduced state as
\begin{equation}
\gcn{n} = \map(\sensorn{n}) = \nonlinearmap(\sensorn{n},\weights)
\end{equation}
where $\nonlinearmap(\cdot,\weights)$ with $\nonlinearmap: \RR^{\nsensors} \rightarrow \RR^{\nmodes}$ denotes the nonlinear function parametrized by a set of parameters $\weights$. For various complex fluid flow problems, the nonlinear relationship between $\sensorn{n}$ and $\gcn{n}$ is rarely obvious. Therefore, in this work, we propose to model $\nonlinearmap(\sensorn{n},\weights)$ via a more general approach using neural networks. We refer to this proposed approach as deep state estimation (DSE).


\subsection{Neural network architecture}
\label{neuralnet}

In this work, we employ a neural network architecture consisting of $\nlayers$ fully-connected layers represented as a composition of $\nlayers$ functions,
\begin{equation}
\nonlinearmap(\dummyinput; \weights) = \hiddenlayer{\nlayers}(\cdot; \weightmat{\nlayers}) \circ \ldots \circ \hiddenlayer{2}(\cdot; \weightmat{2}) \circ \hiddenlayer{1}(\dummyinput; \weightmat{1})
\end{equation}
where the output vector at the $i^{th}$ layer ($i=1,2,\ldots,\nlayers$) is  given by $\hiddenlayer{i}(\dummyinput; \weightmat{i}) = \activation(\weightmat{i}[\dummyinput^T,1]^T)$ with $\hiddenlayer{i}(\cdot; \weightmat{i}): \RR^{\layersize{i-1}} \rightarrow \RR^{\layersize{i}}$. Essentially, an affine transformation of the input vector followed by a point-wise evaluation of a nonlinear activation function, $\activation(\cdot): \RR \rightarrow \RR$, is performed. Here, $\layersize{i}$ is the size of the output at layer $i$ and $\weightmat{i} \in \RR^{\layersize{i} \times \layersize{i-1}+1}$ comprises the weights and biases corresponding to layer $i$.


%
%

A schematic of our proposed DSE approach exhibiting this neural network architecture with $\nlayers=3$ fully connected layers is shown in Fig. \ref{schematic}. Sensor measurements of dimensions $\layersize{0}=\nsensors$ are fed as inputs while the output layer comprising of the reduced state has dimensions $\layersize{\nlayers}=\nmodes$. We note that other architectures such as graph convolutional or recurrent neural networks for exploiting spatial or temporal locality of sensor measurements, respectively, could be utilized to construct the neural network. However, we choose fully-connected layers owing to its simplicity and small dimensions of input and output. 
The weights $\weights \equiv (\weightmat{1},\ldots,\weightmat{\nlayers})$ are evaluated by training the neural network, the details of which are provided in the next section.


\subsection{Training the neural network}
\label{neuralnettraining}


The first step in training is to collect snapshots of high-dimensional states, sensor measurements and reduced states. The FOM Eq. \eqref{fom} is solved at $\nparams$ sampled parameters $\parametersampled$ and $\ntimes$ time instants to obtain the snapshots $\staten{n}(\parametersampledn{i})$ for $i=1,\ldots,\nparams$ and $n=1,\ldots,\ntimes$ resulting in a total of $\nsnapshots=\ntimes\nparams$ snapshots. Then, $\sensorn{n}(\parametersampledn{i})$ is evaluated via some known transformation of $\staten{n}(\parametersampledn{i})$ and $\gcn{n}(\parametersampledn{i})$ is derived by solving
\begin{equation}
\gcn{n}(\parametersampledn{i}) = \underset{\gcdummyn{n}}{\text{arg min}}\| \staten{n} -\basis (\gcdummyn{n}) \|_2^2
\label{optimalreducedstate}
\end{equation}
When $\basis$ is linear with orthogonal columns, for instance POD modes, the solution to Eq. \eqref{optimalreducedstate} is obtained by $\gcn{n}(\parametersampledn{i}) = \basis^T \staten{n}(\parametersampledn{i})$.
%
%
Next, the snapshots of $\sensorn{n}(\parametersampledn{i})$ and $\gcn{n}(\parametersampledn{i})$ for $i=1,\ldots,\nparams$ and $n=1,\ldots,\ntimes$ are standardized via z-score normalization \citep{goodfellow2016deep} to enable faster convergence while training the neural network. 

%
%

Typically, prior to training, the data is divided into  training and validation sets which are used to evaluate/update the weights $\weights$ and test the accuracy of the network, respectively. Accordingly, for each sampled parameter, $\parametersampledn{i}$, snapshots at $\ntrain$ and $\nvalid=\ntimes - \ntrain$ random time instants are chosen for training and validation, respectively, resulting in a total of $\ntrainsnapshots=\ntrain\nparams$ training and  $\nvalidsnapshots=\nvalid\nparams$ validation snapshots. 
Once the neural network is trained, it is tested on a set of testing snapshots which were neither utilized in training nor validation. Accordingly, we collect $\ntestsnapshots=\ntest\nparamspred$ testing snapshots of $\staten{n}(\parameterpredn{i})$, $\sensorn{n}(\parameterpredn{i})$, $\gcn{n}(\parameterpredn{i})$ for $i=1,\ldots,\nparamspred$ and $n=1,\ldots,\ntest$ time instants evaluated at $\nparamspred$ unsampled parameters such that $\parameterpred \nsubseteq \parametersampled$.


We train the neural network $\nonlinearmap(\cdot;\weights)$ to evaluate the trainable parameters $\weights$ by minimizing the $\Ltwo$ error between the reduced state and its approximation, given by
\begin{equation}
\weights = \underset{\weightsdummy}{\text{arg min}}\sum_{i=1}^{\ntrainsnapshots}\| \gc^{(i)} - \nonlinearmap(\sensor^{(i)},\weightsdummy)  \|_2^2
\label{eqn:wgthseqn}
\end{equation}
where the superscript $(i)$ denotes the $i^{th}$ training snapshot. The problem \eqref{eqn:wgthseqn} is solved using stochastic gradient descent (SGD) method with mini-batching and early stopping (which acts as a regularizer to avoid overfitting) \citep{goodfellow2016deep}. 


\section{Numerical experiments: flow over flat plate}
\label{experiments}

In this section, our proposed deep state estimation (DSE) approach is applied on a test case of a two-dimensional (2D) flow over a flat plate. We choose $\basis$ to be linear containing the first $\nmodes$ POD modes of the snapshot matrix, whose columns are given by $\staten{n}(\parametersampledn{i})-\statemean$ for $i=1,\ldots,\nparams$ and $n=1,\ldots,\ntimes$, where $\statemean = 1/M\sum_{i=1,n=1}^{i=\nparams,n=\ntimes} \staten{n}(\parametersampledn{i}) \in \RR^{\statesize}$ is the mean. We again emphasize that POD is chosen for its ubiquity in practice and ease of presentation; the estimation framework described above can be incorporated into a range of ROMs.

All results reported in this section are predictive. That is, the estimated states all lie in parameter regions $\parameterpred$ not sampled for training the neural network. The results generated by DSE are compared with gappy-POD and LSE, described in Sec. \ref{prior}. We also compare the results with the \emph{optimal} reduced states obtained by projection, $\gcn{n} = \basis^T \staten{n}$. These are called optimal because the POD coefficients are exact, and all error is incurred from the ROM approximation \eqref{romapprox}. By contrast, ROM-based SE approaches incur error due to both the ROM approximation and the model error associated with $\map$.  Therefore, none of the above-mentioned SE methods can be expected to estimate a more accurate state than the optimal reconstruction.
The performance of these approaches is analyzed by computing the relative error 
\begin{equation}
Error(\%) = \frac{\|\staten{n}(\parameterpred)-\stateapproxn{n}(\parameteronedpred)\|_2}{\|\staten{n}(\parameterpred)\|_2} \times 100
\label{rerror}
\end{equation}
where $\staten{n}(\parameterpred)$ and $\stateapproxn{n}(\parameterpred)$ are the FOM and estimated solutions, respectively.


\subsection{Problem description}
\label{problemdescription}

We consider flow over a flat plate of length $1$ unit at $\reynolds=200$. The parameter of interest $\parameteroned$ is the angle of attack (AoA) of the flat plate.  Two sets of AoA are considered: a. $\parameteroned \in [25^\circ, 27^\circ]$, and b. $\parameteroned \in [70^\circ, 71^\circ]$. Both parameter sets lead to separated flow and vortex shedding. 

The problem is simulated using the incompressible Navier-Stokes equations in the immersed boundary framework of \cite{colonius2008fast}, which utilizes a discrete vorticity-streamfunction formulation.  The solver employs a fast multi-domain approach for handling far field Dirichlet boundary conditions of zero vorticity. Accordingly, for the two sets of AoA considered, we utilize 5 grid levels of increasing coarseness. The grid spacing of the finest domain (and of the flat plate) is $\dx=0.01$, and the time step is $\dt=0.0008$. All the snapshots are collected after roughly five shedding cycles, by which the system approximately reaches limit cycle oscillations of vortex shedding (and lift and drag). The domain sizes of the finest and coarsest grid levels are a. $[-1,4] \times [-2,2]$ and $[-37.5,42.5] \times [-32,32]$; b. $[-1,3] \times [-1,5]$ and $[-30,34] \times [-45,51]$. The total number of grid points in the finest domain is thus a. $500\times 400$ and b. $400\times 600$, respectively. 
All collected snapshots and state estimation results correspond to the finest domain only. 

Note that all simulations are conducted by placing the flat plate at $0^\circ$ and aligning the flow at angle $\parameteroned$ with respect to the plate. This is done to obtain POD modes that are a good low-dimensional representation of the high-dimensional flow-field for the range of AoAs considered. However, while displaying the results, the flow-fields are rotated back to align the plate at angle $\parameteroned$ for readability.



For state estimation via DSE, the neural network consists of $N_l=3$ layers with dimensions $\layersize{i}=500$ for $i=1,2$, $\layersize{0}=\nsensors$ and $\layersize{\nlayers}=\nmodes$. For the nonlinear activation function $\activation$, we use rectified linear units (ReLU) \citep{goodfellow2016deep} at the hidden layers and an identity function at the output layer. Following the data segregation strategy explained in Sec \ref{neuralnettraining}, training data is split into 80\% and 20\% for training and validation, respectively. For SGD, learning rate is set to 0.1 during the first 500 epochs (for faster convergence) which is then reduced to 0.01 for the remainder of training, momentum is set to 0.9 and mini-batch size is set to 80. Training is terminated when the error on validation dataset does not reduce over 100 epochs, which is chosen as the early stopping criteria. Overall, the network is trained for approximately 2000 epochs on Pytorch.


\subsection{Flow at $\parameteroned \in [25^\circ, 27^\circ]$}

Here we compare the predictive capabilities of our proposed DSE approach to several linear SE approaches for AoA $\parameteroned \in [25^\circ, 27^\circ]$, and where the sensors measure vorticity at $\nsensors=5$ locations on the body. The matrix $\basis$ used for the ROM is constructed using $\nmodes=25$ POD modes of the vorticity snapshot matrix. 

The AoAs used for training the state estimation methods are $\parameteronedsampled=\{25^\circ,25.2^\circ,\ldots,27^\circ\}$, and those used for testing are $\parameteronedpred=\{ 25.5^\circ, 26.25^\circ, 26.75^\circ \} \nsubseteq \parameteronedsampled$. For each AoA, $\ntimes=250$ training snapshots and $\ntest=50$ test snapshots are sampled between $t = 20$ and $28$ convective time units. Thus, a total of $\nsnapshots=2750$ training and $\ntestsnapshots=150$ testing snapshots are used, respectively.

\begin{figure}
\centering
\hspace{-0.4cm}
\begin{subfigure}[t]{0.43\textwidth}
\centering
\includegraphics[scale=0.33]{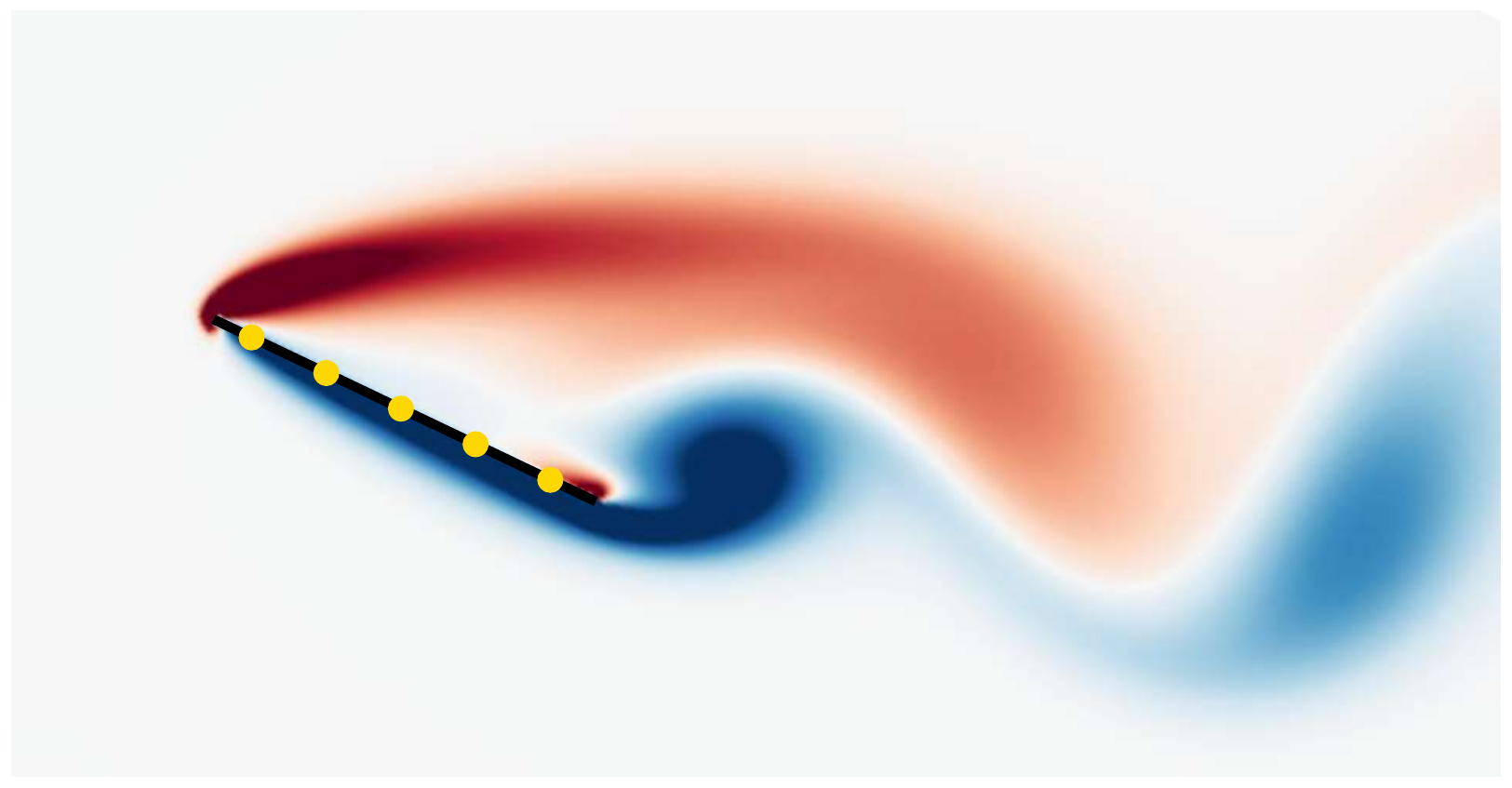}
\caption{\footnotesize  FOM}
\label{demo1fom}
\end{subfigure}
\begin{subfigure}[t]{0.43\textwidth}
\centering
\includegraphics[scale=0.33]{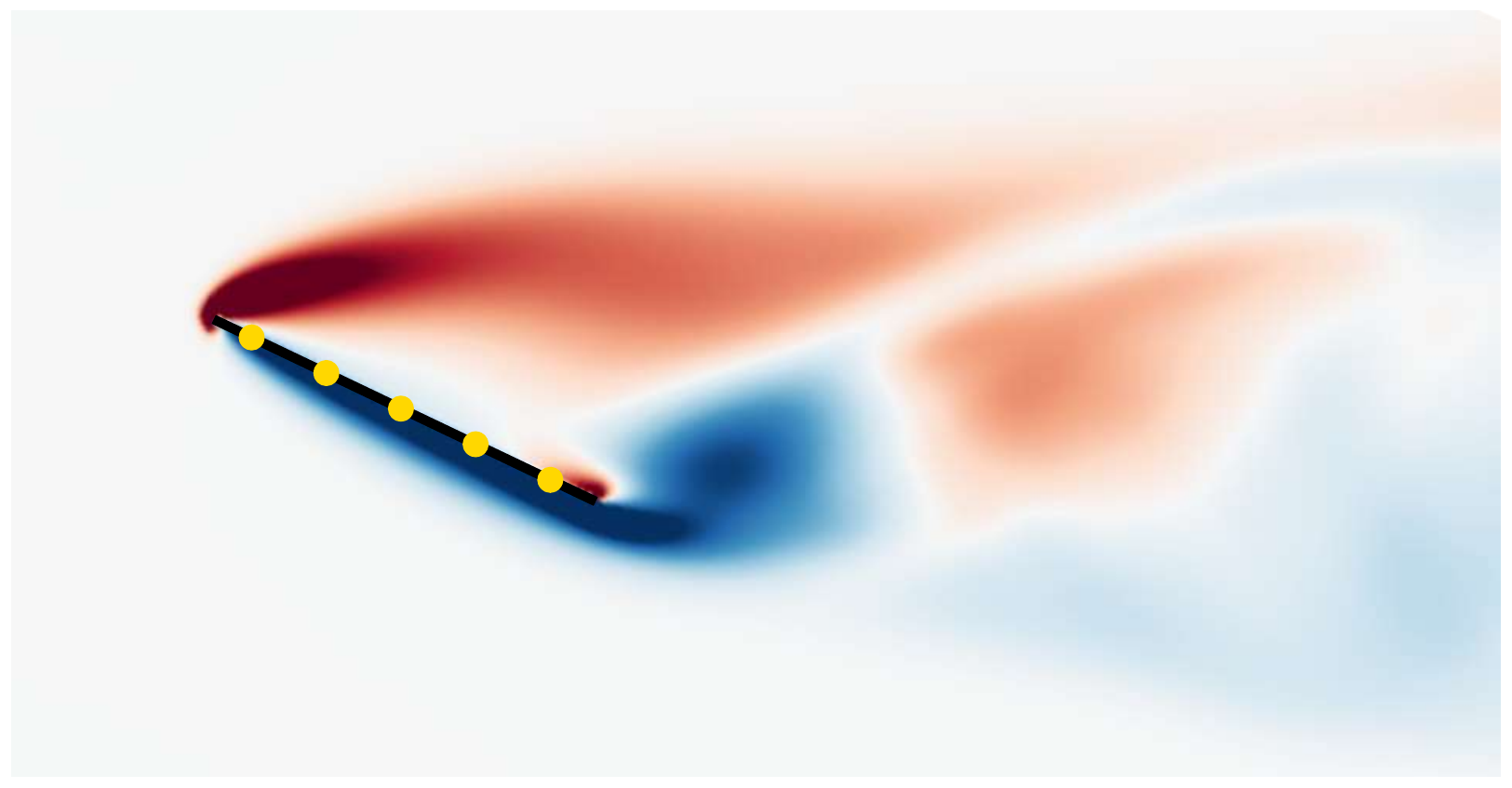}
\caption{\footnotesize Gappy-POD: $34.74 \pm 5.8 \%$}
\label{demo1gpod}
\end{subfigure}
\begin{subfigure}[t]{0.43\textwidth}
\hspace{-0.4cm}
\centering
\includegraphics[scale=0.33]{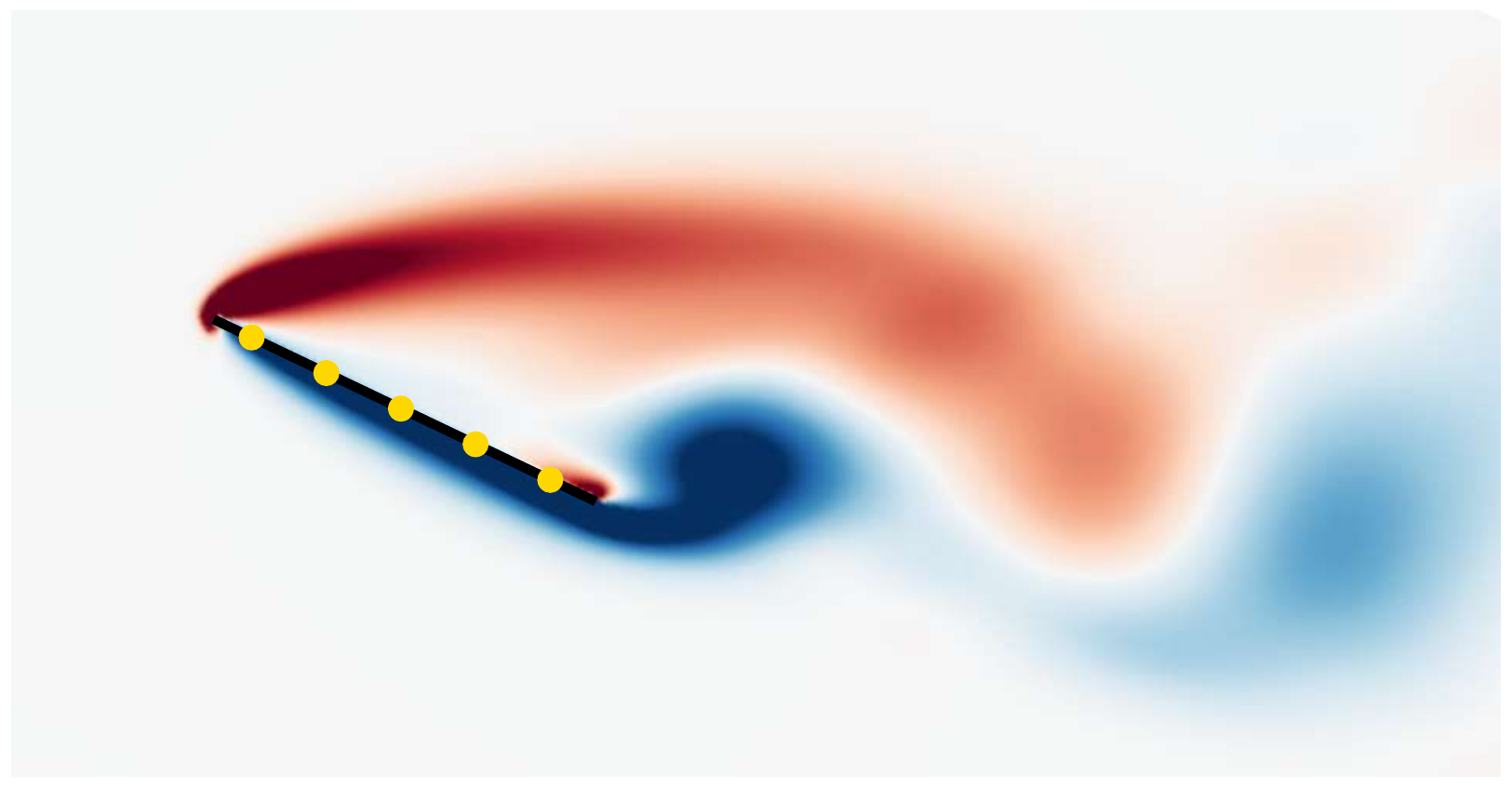}
\caption{\footnotesize  LSE: $16.84 \pm 3.42 \%$}
\label{demo1lse}
\end{subfigure}
\begin{subfigure}[t]{0.43\textwidth}
\hspace{-0.4cm}
\centering
\includegraphics[scale=0.33]{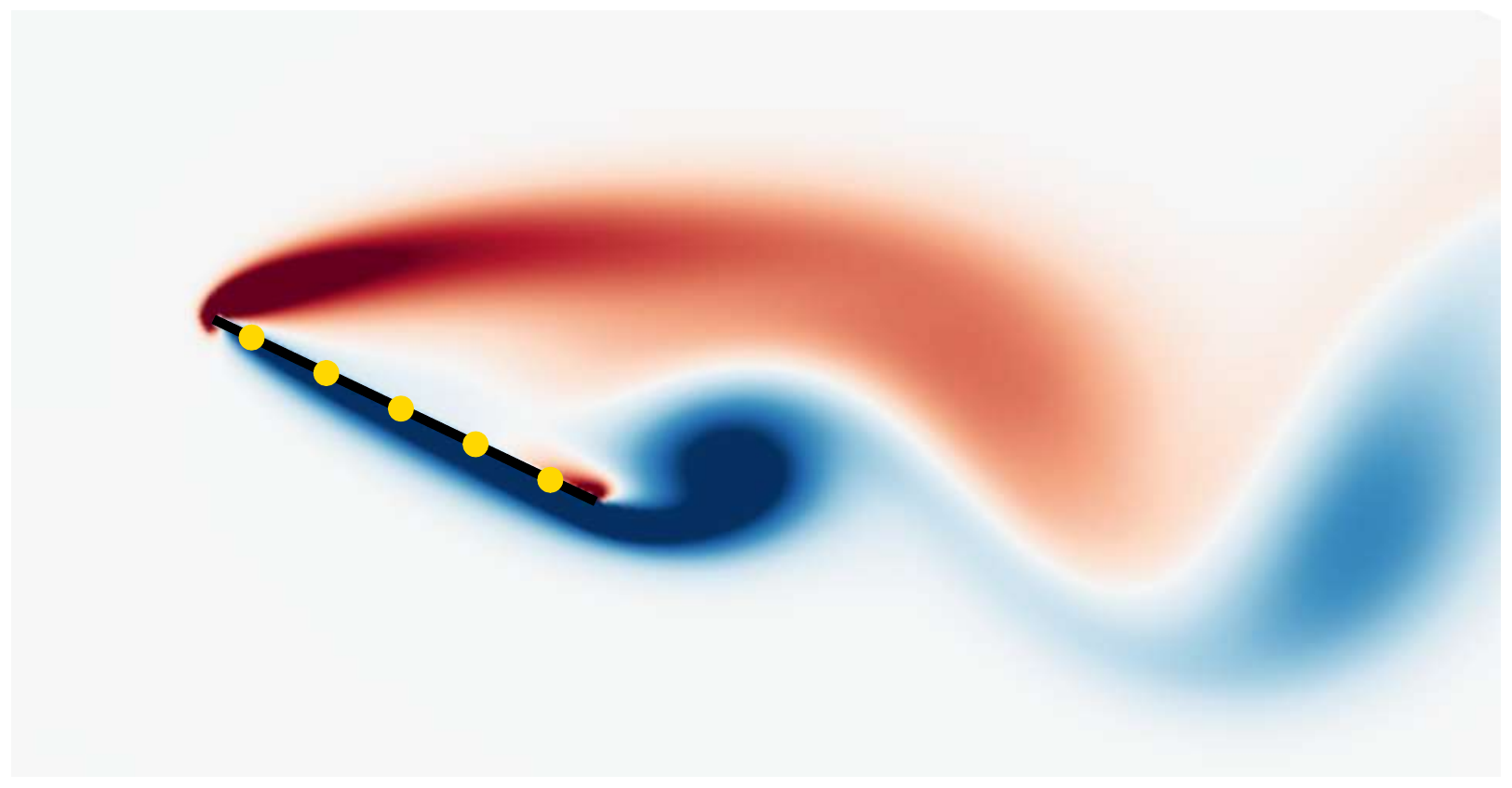}
\caption{\footnotesize DSE: $1.03 \pm 0.25 \%$}
\label{demo1dse}
\end{subfigure}
\caption{\footnotesize Comparison of vorticity contours estimated by the high-fidelity model (FOM), gappy-POD, linear stochastic estimation (LSE), and deep state estimation (DSE) at $\parameteronedpred \in [25^\circ,27^\circ]$ and corresponding relative error with $\nmodes=25$ modes and $\nsensors=5$ sensors that measure vorticity.}
\label{demo1}
\end{figure}

Fig. \ref{demo1} shows the vorticity contours produced by the high-fidelity model (FOM), gappy-POD, linear stochastic estimation (LSE), and deep state estimation (DSE) at a representative instance among the $150$ testing instances. The flow-field constructed by DSE more accurately matches the FOM solution as compared to gappy-POD and LSE. Moreover, the average relative error of the states estimated by DSE is only 1.03\% as compared to 34.74\% and 16.84\% due to gappy-POD and LSE, respectively.

\subsection{Flow at $\parameteroned \in [70^\circ, 71^\circ]$}

\begin{figure}
\centering
\hspace{-0.4cm}
\begin{subfigure}[t]{0.46\textwidth}
\centering
\includegraphics[scale=1]{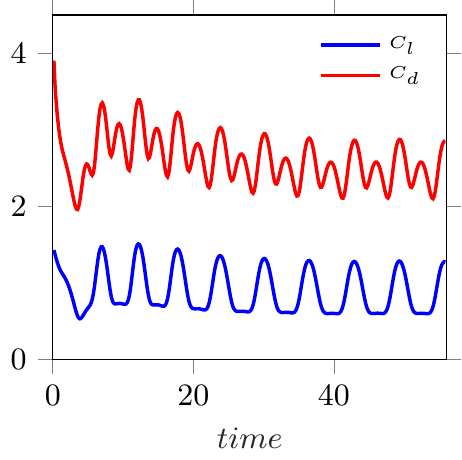}
\caption{Plot of $C_l$ and $C_d$ v/s time at $\parameteroned=70^\circ$}
\label{clcd}
\end{subfigure}
\begin{subfigure}[t]{0.46\textwidth}
\centering
\includegraphics[scale=1]{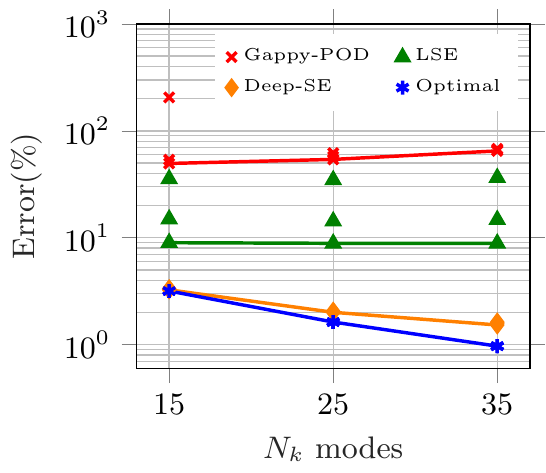}
\caption{\footnotesize Relative error for varying number of surface stress sensors in $\nsensors=\{5,10,15\}$}
\label{errorplot70force}
\end{subfigure}
\caption{\footnotesize Demonstration of intricate vortex shedding (left) and performance comparison of gappy-POD, LSE, DSE and optimal reconstruction when $\parameteronedpred \in [70^\circ,71^\circ]$ (right)}
\end{figure}

We now consider a more strenuous test case of flow at large angle of attack, which exhibits richer dynamics associated with more complex vortex shedding behavior; \emph{c.f.}, Fig. \ref{clcd}. For added complexity and application relevance, we consider sensors that measure the magnitude of surface stress (instead of vorticity) at locations on the body. The basis $\basis$ is constructed from POD modes of the snapshot matrix containing vorticity and surface stress. 
The AoAs for training and testing are $\parameteronedsampled=\{70^\circ,70.2^\circ,\ldots,71^\circ\}$ and $\parameteronedpred=\{ 70.25^\circ, 70.5^\circ, 70.75^\circ \} \nsubseteq \parameteronedsampled$, respectively. For each AoA, $\ntimes=400$ training snapshots and $\ntest=80$ test snapshots are sampled between $t=25$ and 52.2 convective time units, resulting in a total of $\nsnapshots=2400$ and $\ntestsnapshots=240$ snapshots for training and state estimation, respectively. 

\begin{figure}
\centering
\begin{subfigure}[t]{0.46\textwidth}
\centering
\includegraphics[scale=0.28]{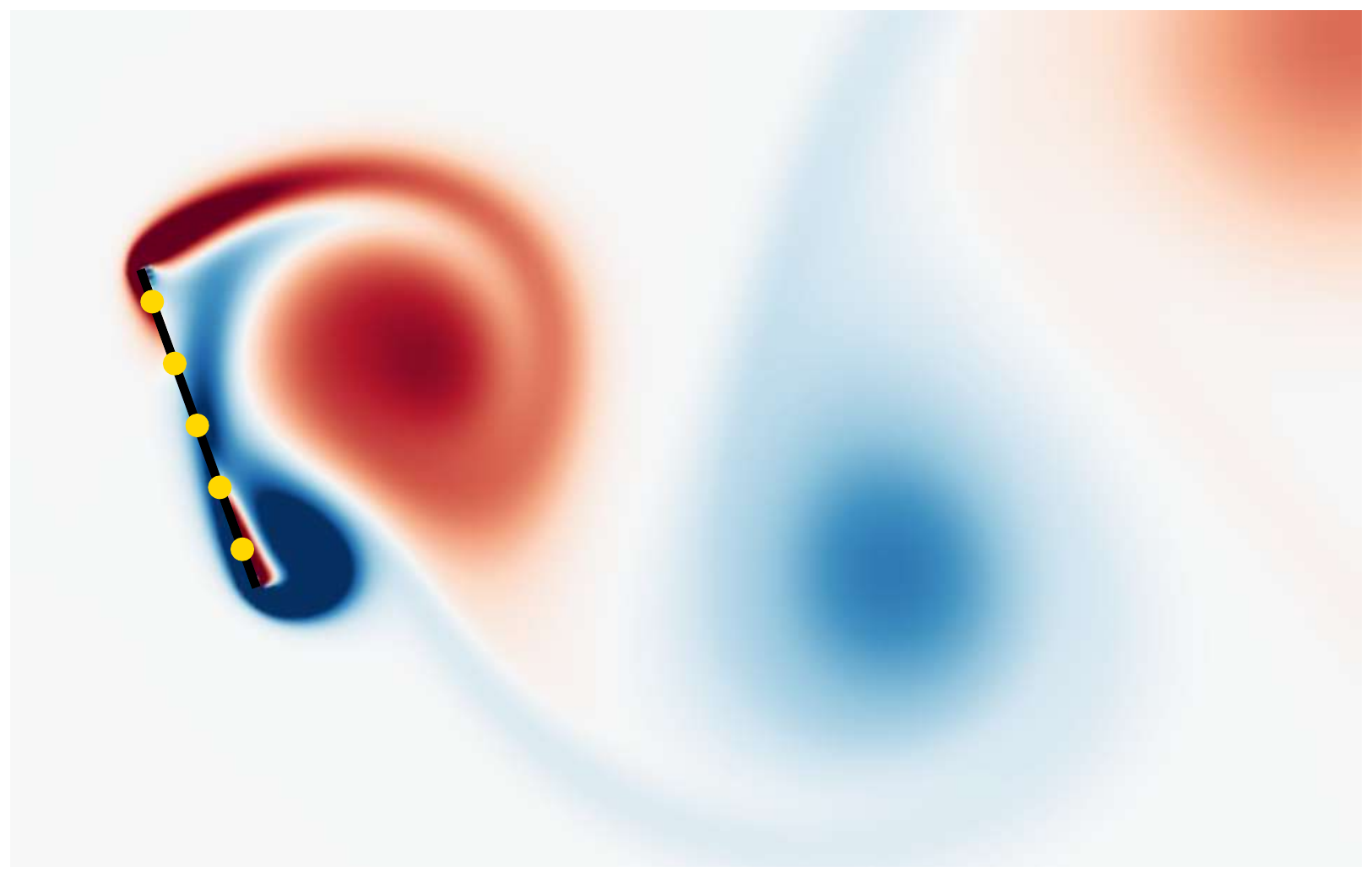}
\caption{\footnotesize  FOM}
\end{subfigure}
\begin{subfigure}[t]{0.46\textwidth}
\centering
\includegraphics[scale=0.28]{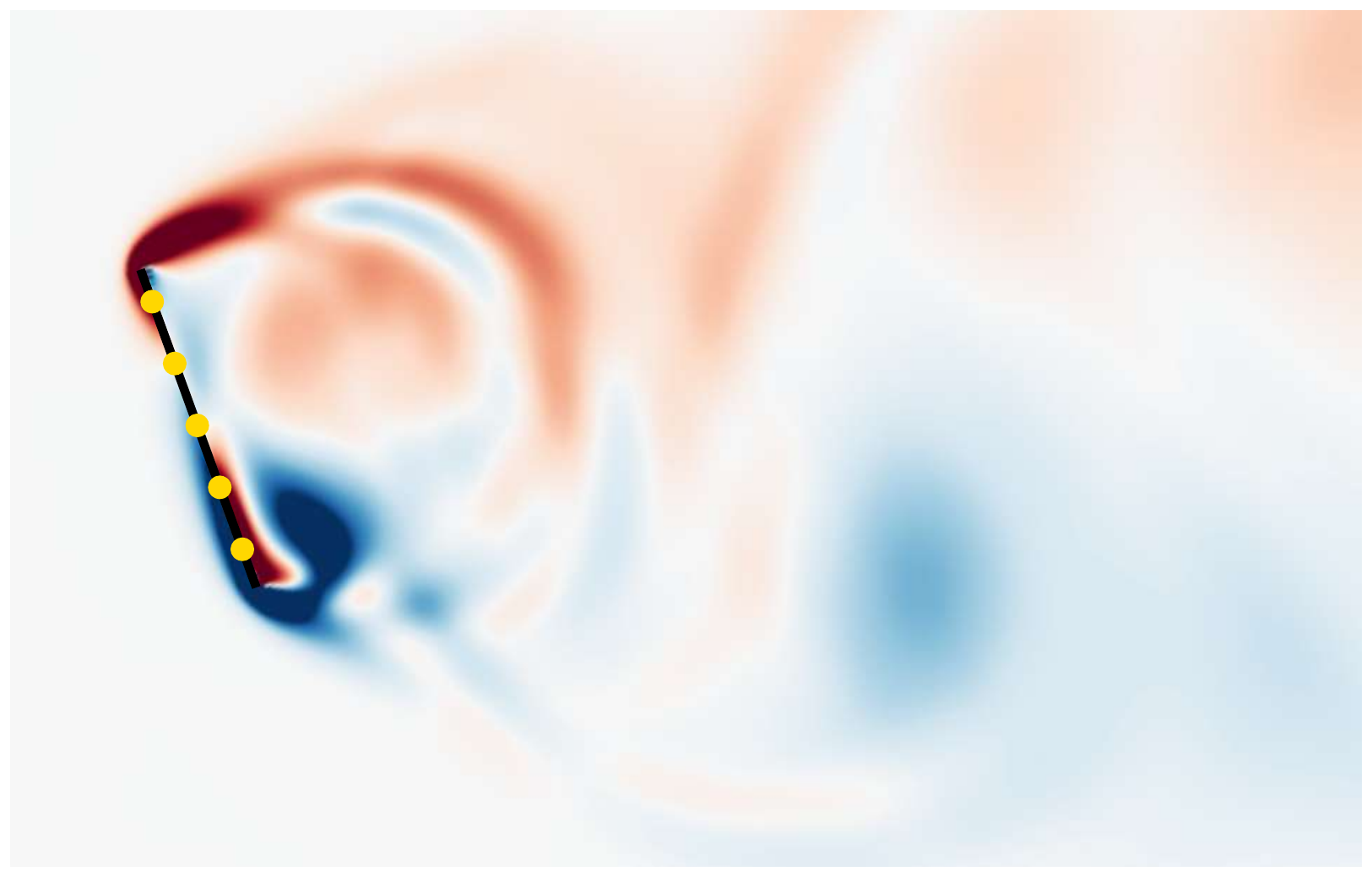}
\caption{\footnotesize Gappy-POD: $61.75 \pm 8.68 \%$}
\end{subfigure}
\centering
\begin{subfigure}[t]{0.46\textwidth}
\centering
\includegraphics[scale=0.28]{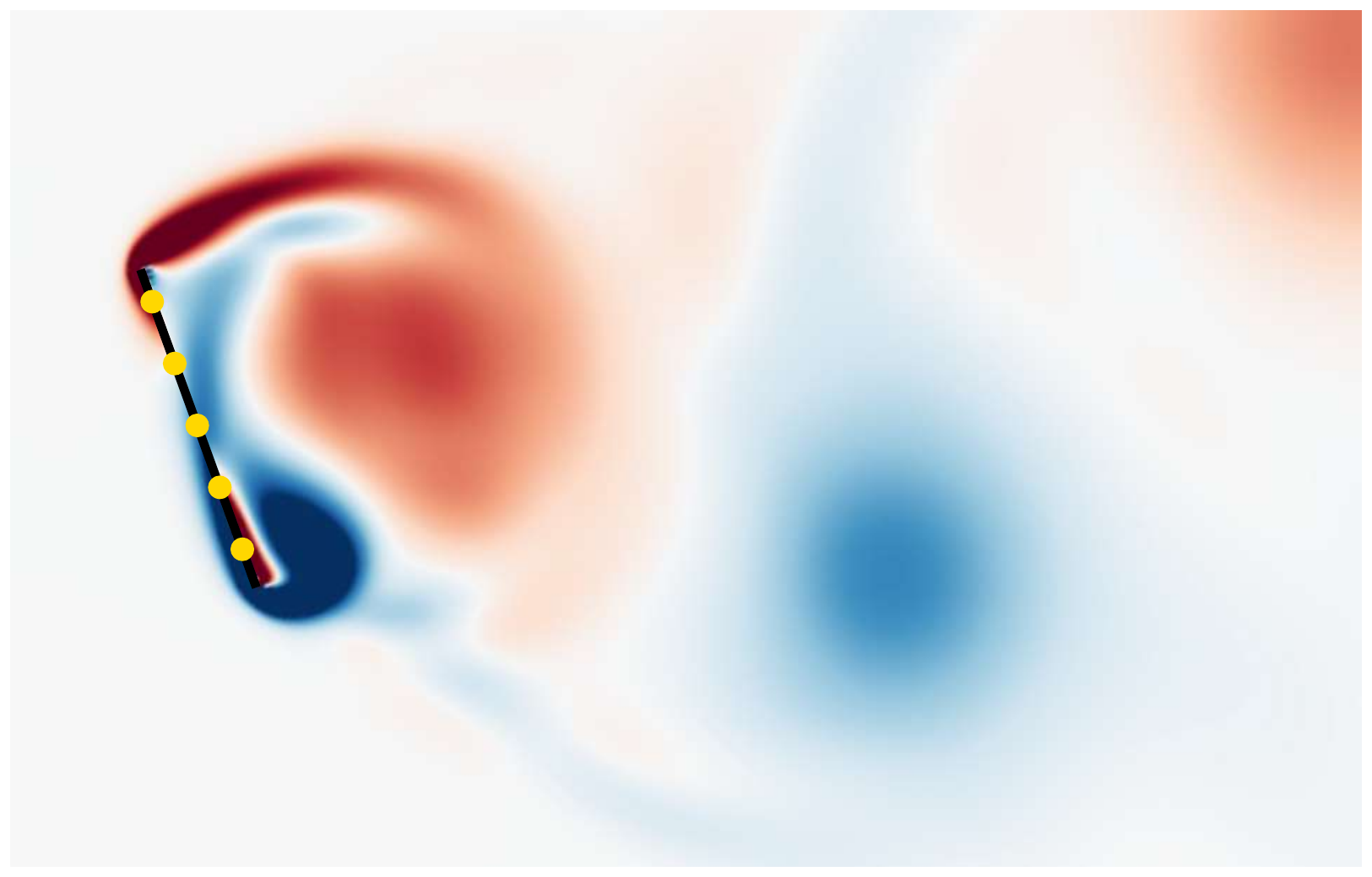}
\caption{\footnotesize  LSE: $35.75 \pm 18.40 \%$}
\end{subfigure}
\begin{subfigure}[t]{0.46\textwidth}
\centering
\includegraphics[scale=0.28]{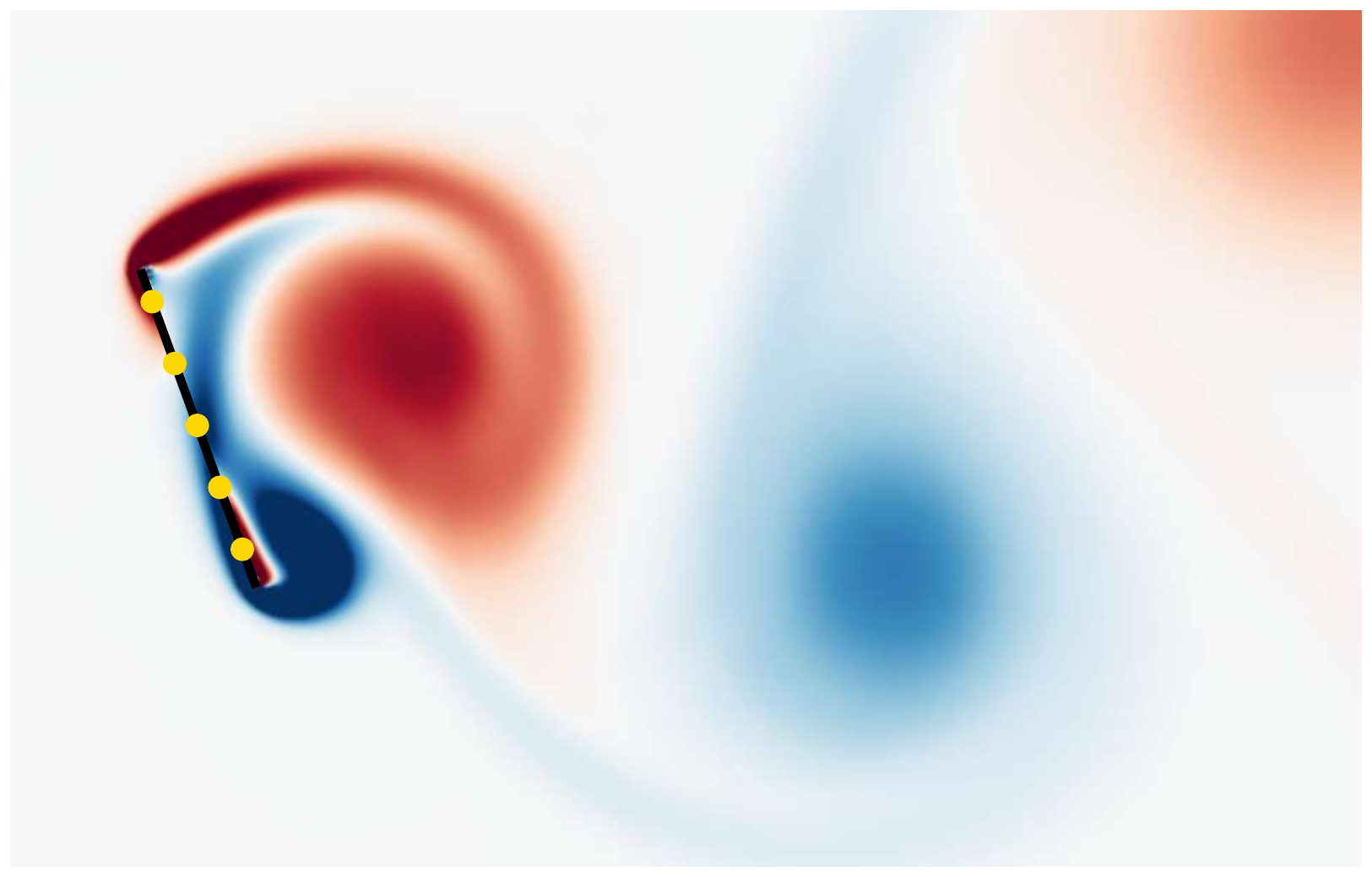}
\caption{\footnotesize DSE: $2.07 \pm 0.38 \%$}
\end{subfigure}
\caption{\footnotesize Analog of figure \ref{demo1} for $\parameteronedpred \in [70^\circ,71^\circ]$, $\nmodes=25$ modes, and $\nsensors=5$ sensors that measure the magnitude of surface stress.}
\label{demo2}
\end{figure}

In Fig. \ref{demo2}, we compare various methods through vorticity contours at a representative instance among the $240$ testing instances obtained by using $N_k = 25$ POD modes and $N_s=5$ sensors. 
It can be observed that DSE significantly outperforms gappy-POD and LSE. Moreover, the average relative error of the states estimated by DSE is only 2.07\% as compared to 61.75\% and 35.75\% in gappy-POD and LSE approaches, respectively.

Finally, we compare the performances of these SE approaches for different numbers of POD modes, $\nmodes=\{15,25,35\}$, and sensors, $\nsensors=\{5,10,15\}$. The average relative errors of the estimated vorticity for these 9 permutations are plotted as markers in Fig. \ref{errorplot70force}. The solid lines connect the markers with lowest errors corresponding to each $\nmodes$, therefore highlighting the best performance among $\nsensors=\{5,10,15\}$. Additionally, the error in \emph{optimal} reconstruction is plotted in blue. Recall that this blue curve only represents a lower bound of the error and is not a SE approach. From the plot, it can be observed that DSE produces an error of only $1-3\%$ as compared to $10-40\%$ due to LSE and $50-200\%$ due to gappy-POD. For all number of sensors considered, DSE produces errors that are comparable to the lower bound. Estimates by LSE do not improve as the number  of modes $\nmodes$ is increased, due to its rank-related limitations as described in Sec. \ref{lsedrawbacks}. 
Similar error trends were also observed for the previous simpler test case of $\parameteroned \in [25^\circ, 27^\circ]$, though  for conciseness these results are not shown in this article.

\section{Conclusions}

In this manuscript, a deep state estimation (DSE) approach was introduced that exploits a low-order representation of the flow-field to seamlessly integrate sensor measurements into reduced-order models (ROMs). In this method, the sensor data and  reduced state are nonlinearly related using neural networks. 
The estimated reduced state can be used as an initial condition to efficiently predict future states or recover the instantaneous full flow-field via the ROM approximation.
Numerical experiments consisted of 2D flow over a flat plate at high angles of attack, resulting in separated flow and associated vortex shedding processes.
At parameter instances not observed during training, DSE was demonstrated to significantly outperform traditional linear estimation approaches such as gappy-POD and linear stochastic estimation (LSE). The robustness of the approach to sensor locations and the physical quantities measured was demonstrated by placing varying number of vorticity- and surface stress-measuring sensors on the body of the flat plate. 
Finally, it is emphasized that the proposed approach is agnostic to the ROM employed; \emph{i.e.}, while a POD-based ROM was utilized for the numerical experiments, the general DSE framework allows for any choice of linear or nonlinear low-dimensional representation.


\section{Declaration of interests}

The authors report no conflict of interest.

\bibliographystyle{jfm}
\bibliography{references}

\end{document}